\def\nodata{ ~$\cdots$~ }
\documentclass{aa}
\usepackage{epsfig}
\usepackage{epsf}
\usepackage{rotate}
\usepackage{rotating}
\begin{document}

   \thesaurus{08(08.06.2; 09.03.1)}

\title{{\it ROSAT} PSPC observations of T Tauri stars in MBM12\thanks{Table~4
is only available in electronic form at the CDS via anonymous ftp to
cdsarc.u-strasbg.fr (130.79.128.5) or via
http://cdsweb.u-strasbg.fr/Abstract.html.}}


   \author{T. Hearty\inst{1}
           \and
           R. Neuh\"auser\inst{1}
           \and
           B. Stelzer\inst{1}
           \and
           M.~Fern\'andez\inst{1}
           \and
           J.M.~Alcal\'a\inst{2}
           \and
           E. Covino\inst{2}
           \and
           V. Hambaryan\inst{3}
          }

   \offprints{thearty@xray.mpe.mpg.de}

   \institute{Max-Planck-Inst. f\"ur Extraterrestrische Physik,
              D-85740 Garching, Germany
          \and
              Osservatorio Astronomico di Capodimonte, I-80131 Napoli, Italy
          \and
              Astrophysikalisches Institut Potsdam, An der Sternwarte 16, D-14482, Potsdam, Germany}

   \date{Received 25 May, 1999; accepted 22 October, 1999}

   \maketitle

   \begin{abstract}
We present the {\it ROSAT} PSPC pointed and {\it ROSAT} All-Sky
Survey (RASS) observations and the results of our low and high spectral
resolution optical follow-up observations of the T~Tauri stars (TTS) and
X-ray selected T~Tauri star candidates in the region of 
the high galactic latitude dark cloud MBM12 (L1453-L1454, L1457, L1458).
Previous observations have revealed 3 ``classical'' T~Tauri stars and 1
``weak-line'' T~Tauri star along the line of sight to the cloud.
Because of the proximity of the cloud to the sun, all of
the previously known TTS along this line of sight were
detected in the 25 ks {\it ROSAT} PSPC pointed observation of the
cloud.  We conducted follow-up optical spectroscopy at the 2.2-meter
telescope at Calar Alto to look for signatures of youth
in additional X-ray selected T~Tauri star
candidates.  These observations allowed us to confirm the existence
of 4 additional TTS associated with the cloud and at least 2 young main
sequence stars that are not associated with the cloud and place an
upper limit on the age of the TTS in MBM12 $\sim$ 10~Myr.

The distance to MBM12 has been revised from the previous
estimate of $65\pm5$~pc to $65\pm35$~pc based on results of the {\it Hipparcos}
satellite.  At this distance MBM12 is the nearest known molecular
cloud to the sun with recent star formation.  We estimate a star-formation
efficiency for the cloud of 2--24\%.

We have also identified a reddened G9 star behind the cloud with
$A_{\rm v}$ $\sim$ 8.4--8.9 mag.  Therefore, there are at least two lines of
sight through the cloud that show larger
extinctions ($A_{\rm v}$ $>$ 5~mag) than previously thought for this cloud. 
This higher extinction explains why MBM12 is capable of star-formation
while most other high-latitude clouds are not.

      \keywords{Stars:formation --
                ISM:clouds
               }
   \end{abstract}

%

\section{Introduction}

The nearest molecular cloud complex to the sun (distance $\sim$ 65~pc)
consists of clouds 11, 12, and 13 from the catalog of
Magnani et al.~(1985) and is  located at (l,b) $\sim$ (159.4,$-$34.3).
This complex of clouds (which we will refer to as MBM12) was first
identified by Lynds (1962) and appears as objects
L1453-L1454, L1457, L1458 in her catalog of dark nebulae.
The mass of the entire complex is estimated
to be $\sim$~30--200~M$_{\odot}$ based on radio maps of the region
in $^{12}$CO, $^{13}$CO and C$^{18}$O (Pound et al.~1990;
Zimmermann \& Ungerechts 1990).

Recently, there has been much interest in understanding the origin
of many isolated T~Tauri stars (TTS) and isolated regions of
star-formation.  For example, within $\sim$~100~pc from the sun there
are at least two additional regions of recent star-formation:
the TW~Hydrae association (distance $\sim$~50~pc; e.g, Kastner et al. 1997;
Webb et al. 1999) and the $\eta$ Chamaeleontis region
(distance $\sim$~97~pc; Mamajek et al. 1999).
Both of these star-forming regions appear to be isolated
in that they do not appear to be associated with any molecular gas.
In addition, both are comprised
mainly of ``weak-line'' TTS\footnote{We define ``weak-line'' TTS (WTTS) to
be TTS with H$\alpha$ equivalent widths, W(H$\alpha$) $>$ $-$10\AA\ and
``classical'' TTS (CTTS) to be TTS with W(H$\alpha$) $<$ $-$10\AA\,
where the negative sign denotes emission}.  In contrast,
most of the TTS in MBM12 are CTTS which are still
associated with their parent molecular cloud.

In addition to the above isolated star-forming regions, TTS have been
found outside of the central cloud core regions in many nearby star-forming
cloud complexes (see references in Feigelson 1996).  Several theories exist
to explain how TTS can separate from their parent molecular clouds either
by dynamical
interactions (Sterzik \& Durisen 1995) or by high-velocity cloud
impacts (\cite{lep94}). Feigelson (1996) also suggests that some of
these TTS may form in small turbulent cloudlets that dissipate after
forming a few TTS.
Since the TTS in MBM12 appear to still be in the cloud in which they formed,
we know they have not been ejected from some other more distant
star-forming region.  Therefore MBM12 may be an example of one of the
cloudlets proposed by Feigelson (1996).
Moriarity-Schieven et al. (1997) argue that MBM12
has recently been compressed by a shock associated with its interface with
the Local Bubble.  This shock may also have recently
triggered the star-formation currently observed in MBM12 (cf. Elmegreen 1993).
Alternatively Ballesteros-Paredes et al. (1999) suggest that
MBM12 may be an example of a star-forming molecular cloud that formed
via large scale streams in the interstellar medium.

MBM12 is different from most other high-latitude clouds at
$\vert b \vert$ $>$ 30$^{\circ}$ in terms of its higher extinction
and its star formation capability (e.g., Hearty et al.~1999).
Based on CO observations and star counts, the peak extinction in the
cloud is $A_{\rm v}$~$\sim$~5~mag (Duerr \& Craine 1982a; Magnani et al.~ 1985;
Pound et al.~1990; Zimmermann \& Ungerechts 1990).
However, molecular clouds are clumpy and it is possible that some
small dense cores with $A_{\rm v}$~$>$~5~mag were not resolved in
previous molecular line and extinction surveys.
For example, Zuckerman et al. (1992) estimate $A_{\rm v}$~$\sim$~11.5~mag
through the cloud, along the line of sight to the
eclipsing cataclysmic variable H0253+193 located behind the cloud
and we estimate $A_{\rm v}$~$\sim$~8.4--8.9 along the line of sight to
a G9 star located on the far side of the cloud (Sect.~\ref{cafos})

Although there is evidence for gravitationally
bound cores in MBM12, the entire cloud does not seem to be bound by
gravity or pressure (Pound et al.~1990; Zimmermann \& Ungerechts 1990).
Therefore, the cloud is likely a short-lived,
transient, object similar to other high latitude clouds which have
estimated lifetimes of a few million years based on the
sound crossing time of the clouds (\cite{mag85}).  If this is the case,
MBM12 will dissipate in a few million years and leave behind an
association similar to the TW~Hydrae association that does not appear
to be associated with any molecular material.

\begin{figure}
\resizebox{\hsize}{!}{\includegraphics{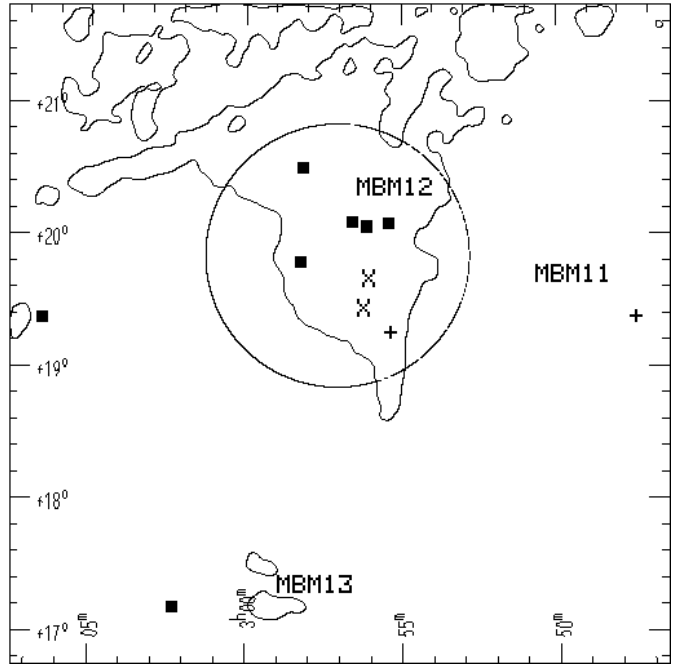}}
\caption{The contour shows the extent of the IRAS 100$\mu$m emission
from MBM12 in the region where we investigated the RASS data.
The circle shows the region observed in the {\it ROSAT} PSPC pointed
observation.  The squares mark the locations of the 8 TTS in MBM12.
The two stars LkH$\alpha$262 and LkH$\alpha$263 are separated by
$\sim$ 15$\arcsec$ and thus not resolved in this figure.
The plus symbols mark the locations of two young {\it ROSAT}
detected main sequence stars which still contain lithium, and the
crosses mark the two lines of sight through the cloud which are known
to have $A_{\rm v}$ $>$ 5 mag.  This and all subsequent figures use J2000
coordinates.}
\label{iras}
\end{figure}

Previous searches for TTS in MBM12 have made use of H$\alpha$,
infrared, and X-ray observations.  
The previously known TTS in MBM12 are listed in Table~\ref{previous}
with their coordinates, spectral types, apparent magnitudes, and selected
references.
We include the star S18 in the list even though Downes \& Keyes (1988)
point out that it could be an Me star rather than a T~Tauri star since our
observations
confirm that it is a CTTS.  The previously known and new TTS stars
identified in this study are plotted in Fig.~\ref{iras} with
an IRAS 100~$\mu$m contour that shows the extent of the cloud.

\begin{table}
\caption{Previously known T~Tauri stars in MBM12}
\label{previous}
\begin{tabular}{@{}cccccc@{}}
\hline
Star & RA & Dec & SpT & $V$  & Ref. \\
     & [2000] & [2000] & & [mag] & \\
\hline
LkH$\alpha$262 & 2:56:07.9 & 20:03:25 & M0 & 14.6     & 1,2    \\
LkH$\alpha$263 & 2:56:08.4 & 20:03:39 & M4 & 14.6     & 1,2    \\
LkH$\alpha$264 & 2:56:37.5 & 20:05:38 & K5 & 12.5     & 1,2,3,4   \\
E02553+2018     & 2:58:11.2 & 20:30:04 & K4 & 10:      & 2,5       \\
S18            & 3:02:20.0 & 17:10:35 & M3 & 13.5  &  6,7   \\
\hline
\end{tabular}

(1) Herbig \& Bell (1988).
(2) Fern\'andez et al. (1995).
(3) Magnani et al. (1995).
(4) Gameiro et al. (1993).
(5) Caillault et al. (1995).
(6) Stephenson (1986).
(7) Downes \& Keyes (1988) point out this could be an Me star rather than
a T~Tauri star.

\end{table}

Although the TTS LkH$\alpha$264 is a well studied object
because of its extreme CTTS features (i.e., strong H$\alpha$ emission
and infrared excess) and it is known to have strong \ion{Li}{I}
$\lambda$6708~\AA\ absorption (Herbig 1977), there is no measurement
for the equivalent width of the lithium line, W(Li), of this star in the
literature.
The only previously known T Tauri star in this cloud which has a
published measurement of W(Li) is
E02553+2018 which was first identified as a T Tauri star in the
{\it Einstein} Extended Medium Sensitivity Survey (EEMSS)
(Gioia et al. 1990; Stocke et al. 1991).  More recently, Mart\'\i n
et al. (1994) presented a high resolution spectrum of
this star for which they measured W[Li] = 475~m\AA. 
Caillault et al.~(1995) reported an X-ray count rate for this source of
0.008 counts s$^{-1}$ in the {\it Einstein} band.  None of the other
TTS in the cloud have been previously detected in X-rays.

We present the {\it Hipparcos} satellite observations
in Sect.~\ref{hipsec} that show the distance to MBM12 is not as well
constrained as previously thought and our high resolution observations
of two TTS in the cloud that indicate the TTS are probably at the same
distance as the cloud.  In Sect.~\ref{rassobs} we describe the
{\it ROSAT} All-Sky Survey (RASS) and {\it ROSAT}
pointed observations investigated.  In Sect.~\ref{cafos} we present
the results of our low-resolution spectroscopic observations of
the T~Tauri star candidates.  In Sect.~\ref{xrayvar} we discuss
the X-ray variability of the TTS and in Sect.~\ref{xlf}
we derive an X-ray luminosity function for the TTS in MBM12.
In Sect.~\ref{conclusions} we present our conclusions.

\section{The distance to MBM12 and its T~Tauri stars}
\label{hipsec}

\subsection{The distance to the gas}

The distance to MBM12 was first estimated by Duerr \& Craine (1982a)
using Wolff diagrams.  They found evidence for two clouds along this line
of sight: one cloud at a distance of 200--300~pc with a typical visual
extinction of $\sim$~2--3 mag and another cloud at a distance of
500--800~pc with a typical visual extinction of $\sim$~1.5~mag.
However, more recent observations show that this is probably one complex
of molecular gas at a much smaller distance.  

A more accurate estimate of the distance was reported by Hobbs et al.~(1986)
and Hobbs et al. (1988).
The method used was to identify 10 bright stars in the direction
of the cloud for which a {\it spectroscopic} parallax could be determined
and look for interstellar NaI~D lines in the spectra of these
stars.  They found that the star HD18404 (distance $\sim$\ 60~pc)
showed no interstellar absorption features and is therefore presumably
in front of the cloud  and the star
HD18519/20 (distance $\sim$\ 70~pc) did show interstellar absorption features
and is therefore behind the cloud.
Since observations of the other 8 stars are consistent
with these two stars, most investigators have assumed a distance
of $\sim$~$65\pm5$~pc for the cloud.  

Since the {\it Hipparcos} satellite measured the {\it trigonometric} parallax
of
most of these stars it is no longer necessary to assume a spectral type or
intrinsic luminosity (as is necessary for a spectroscopic parallax) to
measure their distance.
According to {\it Hipparcos}, the distance to HD18404 is
$\sim$~32$\pm$1~pc and the distance to HD18519/20 is $\sim$~90$\pm$12~pc.
The stars used by Hobbs et al.~(1986) and Hobbs et al. (1988)
to establish the distance to MBM12 are listed in Table~2
along with their apparent magnitude, spectral type, distance based on
spectroscopic parallax, distance based on the {\it Hipparcos} parallax,
and whether the spectrum presented in Hobbs et al.~(1986) and Hobbs et al.
(1988) showed interstellar \ion{Na}{I} absorption lines.
Although the {\it Hipparcos} results indicate the distance to MBM12
is not as well constrained as thought, it is consistent with the
previous result (i.e., $\sim$~65$\pm$35~pc).
However, since the revised distance estimate has $\sim$~50\% uncertainty
it cannot be used to derive accurate parameters for the TTS in the cloud.

\begin{table}
\label{distance}
\caption{Stars from Hobbs et al.~(1986) and Hobbs et al. (1988)}
\begin{tabular}{@{}c@{\hspace{5pt}}c@{\hspace{5pt}}c@{\hspace{5pt}}c@{\hspace{5pt}}c@{\hspace{5pt}}c@{\hspace{5pt}}c@{}}
\hline
HD       & HIP      & $V$  & SpT   & \multicolumn{2}{c}{Distance [pc]}  & NaI  \\
\cline{5-6}
         &          &[mag] &       & spectroscopic & {\it Hipparcos} &      \\
\hline					                        	         
18090    & \nodata  & 8.85 & F3V   & 145   & \nodata    & yes  \\
18091    & 13579    & 7.00 & A9V   & 85    & $96\pm11$  & no   \\
18190    & \nodata  & 8.98 & A9V   & 185   & \nodata    & yes  \\
18256    & 13702    & 5.63 & F6V   & 25    & $35\pm1$   & no   \\
18283    & 13723    & 8.78 & A8III & 380   & $150\pm29$ & yes  \\
18404    & 13834    & 5.80 & F5IV  & 60    & $32\pm1$   & no   \\
18484    & 13892    & 6.70 & A3III & 211   & $141\pm35$ & yes  \\
18508    & 13913    & 7.34 & F2V   & 80    & $91\pm8$   & no   \\
18519/20 & 13914    & 4.63 & A2V   & 70    & $90\pm12$  & yes  \\
18654    & 14021    & 6.79 & A0V   & 160   & $128\pm18$ & yes  \\
\hline
\end{tabular}
\end{table}

Zimmermann \&
Ungerechts (1990) point out that, although the distance estimate based on
the two bright stars from \cite{hob86} and \cite{hob88} is valid for the
northern section of the
cloud, radio observations of the molecular gas show
there are at least 4 velocity components that may all be at
different distances. 
However, the polarization map of MBM12 produced by Bhatt \& Jain (1992)
shows that the
polarization of the stars in the upper region of the cloud
(where the Hobbs et al.~1986 and Hobbs et al.~1988 stars
are located) is the same at that
of the lower region of the cloud.  Thus, they argue that the entire complex
is at the same distance.

Zimmermann \& Ungerechts (1990) also find that the ratio of the
CO mass to the virial mass M(CO)/M$_{\rm vir}$ = 0.03 for the whole cloud
indicating the entire cloud is not gravitationally bound.  However,
they point out that
some of the cores in the larger clumps may be gravitationally bound.
If the cloud is found to be at a somewhat greater distance, these results
will support the hypothesis that a few of the large clumps are
gravitationally bound.

\subsection{The association of the stars with the gas}

Although the TTS and the cloud are projected along the same
line of sight, they may be at different locations and thus
not associated with each other.  However, we can check whether the radial
velocities of the stars are similar with the gas to find out if it is
likely that the T~Tauri stars are associated with the cloud.  The
distribution of cloud velocities are quite large
with at least 4 components with average velocities $-$5.0~km~s$^{-1}$ (I),
$-$2.3~km~s$^{-1}$ (II), 1.4~km~s$^{-1}$ (III), and 5.0~km~s$^{-1}$ (IV)
(Zimmermann \& Ungerechts 1990; Pound et al. 1990;
Moriarty-Schieven et al.~1997).

\begin{table}
\caption{Radial Velocities of the TTS in MBM12}
\label{velocity}
\begin{tabular}{cccc}
\hline
       Star       &      V$_{\rm lsr}$       &     vsin$i$   \\
                  &      [km~s$^{-1}$]   &    [km~s$^{-1}$]  \\
\hline
  RXJ0255.4+2005      & $4.5\pm1.0$       &  $10.0\pm3.0$  \\
  LkH$\alpha264^{a}$  & $-4.2\pm2.5$  &  $24.3\pm2.0$  \\
\hline
\end{tabular}

$^{a}$ Herbig (1977) measured a radial velocity of $\sim~1.0\pm3.9$
km~s$^{-1}$ for this object.

\end{table}

We obtained high resolution spectra of two of the
T~Tauri stars in MBM12 with FOCES at the Calar Alto 2.2-m telescope
in August 1998.  The spectra for these stars
(RXJ0255.4+2005 and LkH$\alpha$264, see Fig.~3.)
allow us to estimate their radial velocities
and confirm the W(Li) measurements
of our low resolution spectra presented in Sect.~\ref{cafos}.
Determinations of radial velocity, RV, and projected rotational velocity,
vsin$i$, have been obtained by means of cross
correlation analysis of the stellar spectra with those of radial velocity
and rotational standard stars, treated in analogous way.
Given the large spectral range covered by the FOCES spectra,
the cross correlation of the target and template stars was performed
after rebinning the spectra to a logarithmic wavelength scale, in order
to eliminate the dependence of Doppler shift on the wavelength.
Moreover, only parts of the spectra free of emission lines and/or not
affected by telluric absorption lines have been used. Therefore, the NaI D,
and H$\alpha$ lines as well as wavelengths longer than about 7000 \AA\ have
been excluded from the cross-correlation analysis.
The result of the cross-correlation is a correlation peak which can be
fitted with a Gaussian curve. The parameters of the Gaussian, center
position and full-width at half-maximum (FWHM) are directly related to 
RV and vsin$i$, respectively.  The method of the correlation has been fully
described by Queloz (1994), and Soderblom et al. (1989). More details about
the calibration procedure can be found in Appendix A of Covino et al. (1997).

The radial velocities we measured for the two MBM12 TTS listed
in Table~\ref{velocity} are similar to that of the molecular gas
(Zimmermann \& Ungerechts 1990; Pound et al. 1990).
Radial velocity measurements have not yet been made for the fainter stars.
Nevertheless, the superposition of the TTS on the
cloud and the similar radial velocities of at least two of the
stars with the gas are strong evidence to support that
the TTS are associated with the cloud.

\section{The {\it ROSAT} observations of MBM12}
\label{rassobs}

\begin{figure*}
\hbox{\epsfxsize=8.9cm \epsfbox{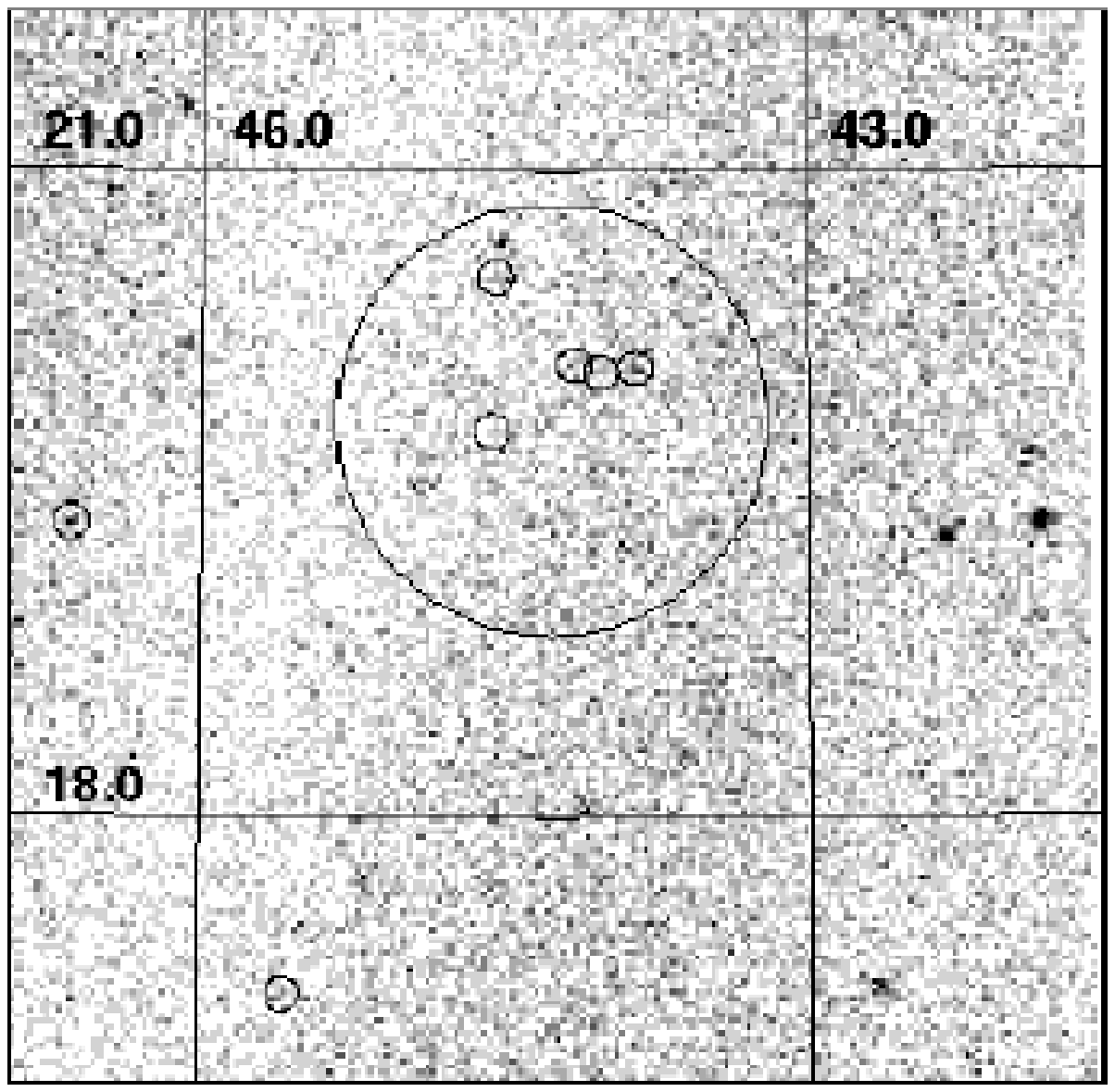}
\epsfxsize=8.9cm \epsfbox{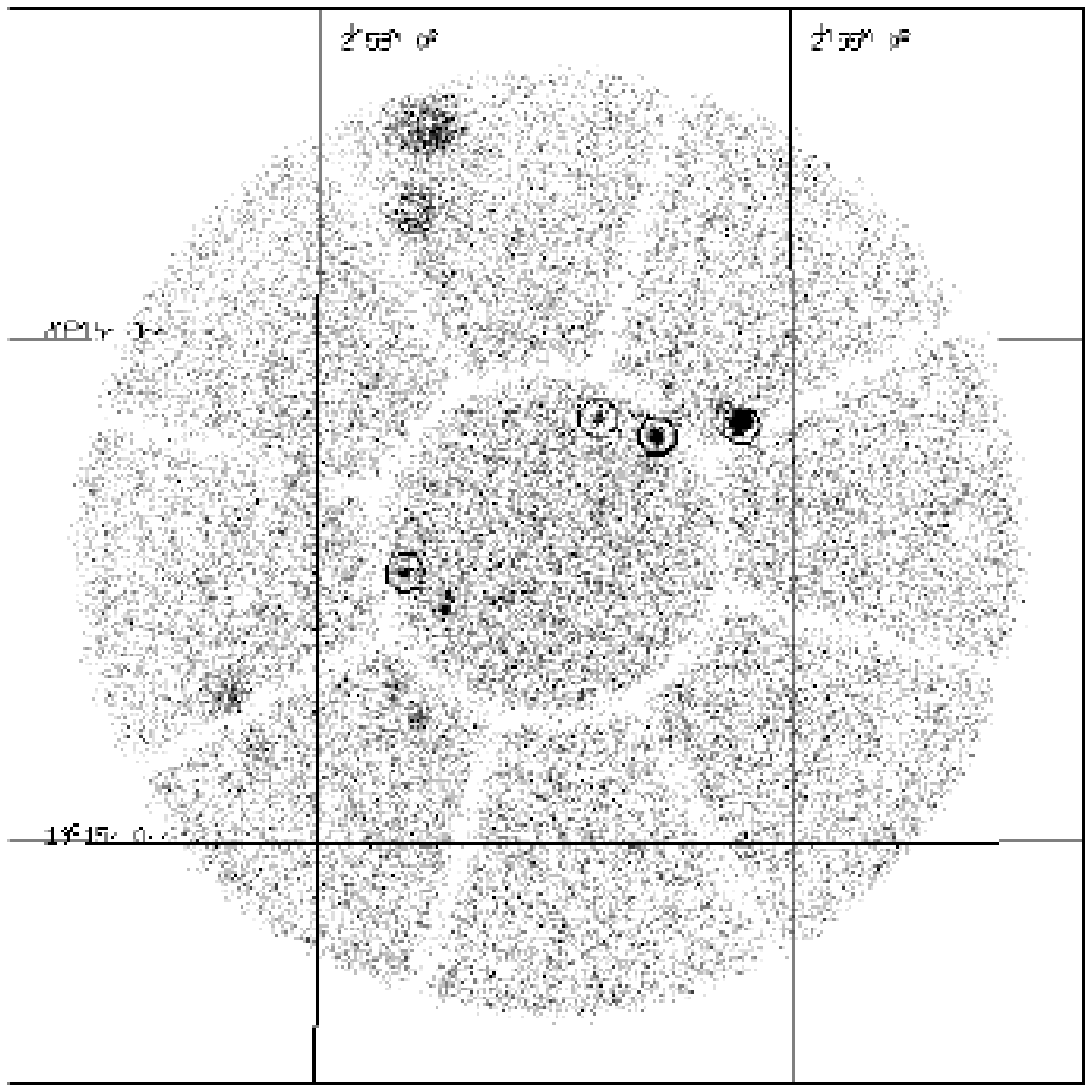}}
\caption{A RASS broad-band (0.08--2.0~keV) image of the region we
investigated around MBM12 is displayed in the left frame.
The large circle showing the field of view of the pointed observation
which is displayed in the right frame.
The small circles mark the locations of the X-ray identified
TTS in MBM12.  The mean RASS exposure time in this region of the sky
is quite low $\sim$~340~s, therefore only 3 of the TTS in this region
were above our detection threshold and one, S18, was slightly below our
threshold.  However all of the TTS in the field of view of the pointed
observation were clearly detected.}
\label{xrayfig}
\end{figure*}

Since both CTTS and WTTS are typically $\sim$~10$^{3}$--10$^{5}$
times more luminous in the X-ray region of the spectrum than
average (i.e., older) low-mass stars (Damiani et al. 1995),
we made use of the {\it ROSAT} pointed and the RASS
observations of MBM12 to identify previously unknown TTS in the cloud.  
The 25~ks {\it ROSAT} PSPC pointed observation (Sequence number 900138) was
centered at (RA,Dec) $\sim$ (2:57:04.8,+19:50:24).  Although they were
originally discovered by other means, all of the previously known TTS in the
central region of MBM12 were also detected with {\it ROSAT}.  
Since the extent of the molecular gas is not known (in particular for the MBM13 region)
and TTS can sometimes be displaced several parsecs from their parent clouds,
we also searched in the RASS database in a $\sim$~25~deg$^2$\ region around
MBM12.  Details about {\it ROSAT}
and its PSPC detector can be found in Tr\"umper (1983) and
Briel \& Pfeffermann (1995), respectively.
The RASS broad-band image of the region investigated around MBM12
and the {\it ROSAT} pointed observation are displayed in Fig.~\ref{xrayfig}.

The X-ray source search was conducted in different
{\it ROSAT} standard ``bands", defined 
as follows: ``broad" = 0.08--2.0~keV; ``soft" = 0.08--0.4~keV;
``hard" = 0.5--2.0~keV; ``hard1" = 0.5--0.9~keV; ``hard2" = 0.9--2.0~keV.
We identified all of the X-ray sources above a maximum likelihood\footnote{The
maximum likelihood can be converted into probability P through the equation
P = 1 $-$ exp($-$ML).} threshold of 7.4 in both the RASS and the {\it ROSAT} PSPC
pointed observations of MBM12.  In addition, we selected only those
X-ray sources above a count rate threshold of $\sim$~0.03 cts~s$^{-1}$\
in the RASS observation and a count rate threshold of
0.0013 cts~s$^{-1}$\ in the {\it ROSAT} PSPC pointed observation.
The one previously known TTS candidate, S18, near the cloud MBM13 was
detected in the RASS with a ML = 6.4 (i.e., below our threshold), however,
since our optical spectroscopic observations confirm that it is a TTS we
include it in our study.
We identified 49 X-ray sources in the {\it ROSAT} PSPC pointed
observation of MBM12 (including all of the
previously known TTS in the central region of the cloud)
and 28 X-ray sources detected in the RASS (including S18) in the
regions displayed in Fig.~\ref{xrayfig}.  Three stars were detected
both on the RASS and in the pointed observation.

We list the sources detected in the {\it ROSAT} PSPC pointed
observation and in the RASS in Table~4.
We include the {\it ROSAT} source name, the X-ray source coordinates,
the maximum likelihood for existence for each source, the broad-band count rates,
the X-ray hardness ratios $HR1$ and $HR2$ (as defined in
Neuh\"auser et al.~1995), the apparent visual magnitude
taken from the Guide Star Catalog (magnitudes for the fainter
sources indicated with a ``:'' are estimated from the digitized
sky survey images), and the broad-band X-ray to optical flux ratio.
We also list the spectral type and the H$\alpha$ and lithium  equivalent
widths of the sources that have been observed spectroscopically and
comments collected from
our search through the SIMBAD\footnote{This research has made use of the
SIMBAD database, operated at CDS, Strasbourg, France.} and NED\footnote{The
NASA/IPAC Extragalactic Database (NED) is operated by the Jet Propulsion
Laboratory, Caltech, under contract with the National Aeronautics
and Space Administration.} databases concerning the objects.

\addtocounter{table}{1}

Assuming a mean X-ray count-rate-to-flux
conversion factor of 1.1~$\times$~10$^{-11}$~erg~cts$^{-1}$~cm$^{-2}$,
which we derive from X-ray spectral fits of the TTS in Sect.~\ref{xlf},
if the cloud is at a distance of 65~pc, the limiting
luminosities of the observations are $1.7 \times 10^{29}$ erg s$^{-1}$
and $7.2 \times 10^{27}$ erg s$^{-1}$ for the RASS and {\it ROSAT}
pointed observations, respectively.  Therefore, these observations are
sufficient to detect most of the WTTS in the cloud since
the threshold is below the X-ray faintest stars in the WTTS
X-ray luminosity function (e.g., Neuh\"auser et al.~1995).  Although the
RASS observation of MBM12 in not sensitive enough to detect
all of the CTTS in the cloud, the objective prism survey by
Stephenson (1986) identified all of the H$\alpha$ emission sources in this
region down to a visual magnitude threshold of $\sim$~13.5.  Since this
limiting magnitude corresponds to the early M spectral types in MBM12,
the current population of TTS in MBM12 presented in this paper
should be complete for all earlier spectral types.

Since MBM12 is at relatively high galactic latitude, many of the 81 X-ray
sources we identified are extragalactic. Therefore, we used the X-ray
to optical flux ratios (see Table~4) to remove extragalactic
sources from our list of candidates (cf. Hearty et al.~1999). 
All sources which have log($f_{\rm x}/f_{\rm v}$)~$>$~0.0
are considered to be extragalactic and those with
log($f_{\rm x}/f_{\rm v}$)~$<$~0.0 are considered to be
stellar objects, some of which could be PMS.  
We also searched the literature to remove cataloged non-PMS stars
from our list of candidates.  Finally we were left with a list of 
X-ray sources identified in the RASS
and {\it ROSAT} pointed observations of the cloud  which have stellar optical
counterparts that may be PMS stars.  However, many of these stars may
be other types of X-ray active stars (e.g., RS CVn and dMe stars)
and nearby main sequence stars (which may not be intrinsically X-ray bright,
but are near enough so that their X-ray flux is large) that
are more difficult to separate from PMS stars by X-ray
observations alone.  Therefore, follow-up spectral observations
are necessary to identify which X-ray sources are T~Tauri stars.

\section{The optical spectroscopy}
\label{cafos}

In order to complete the census of the TTS population of MBM12
we require follow-up observations.  Since
lithium is burned quickly in convective stars, a measurement of
W(Li) along with a knowledge of the spectral
type of a star can be a reliable indicator of youth.  Therefore
we obtained broad-band, low-resolution, optical spectra of the X-ray emitting
TTS candidates to determine spectral types and measure the
equivalent width of the H$\alpha$ emission and \ion{Li}{I}
6708~\AA\ absorption lines.

The spectra were obtained from October 9--11, 1998 with the
Calar Alto Faint Object Spectrograph (CAFOS) at the 2.2-m
telescope at Calar Alto, Spain.  The 24$\mu$m pixels of the SITe-1d
2048$\times$2048
chip with the G-100 grism provided a reciprocal dispersion of
$\sim$~2.1~\AA~pixel$^{-1}$.
The resolving power, $R$ = $\lambda / \delta \lambda$ $\sim$ 1000, derived
from the measurement of the FWHM
(FWHM~$\sim$~6.4~\AA) of several well isolated emission lines of the
comparison spectra is sufficient to resolve the
lithium absorption line in T~Tauri stars.  The wavelength range
$\sim$~4900 to 7800~\AA\ was chosen to detect two indicators of possible
youth (H$\alpha$ emission and \ion{Li}{I}~$\lambda$6708~\AA\ absorption)
and to determine
spectral types. All spectra were given an initial inspection at the telescope.
If a particular star showed signs of youth or the integration
produced fewer than $\sim$~1000 cts pixel$^{-1}$, at least one
additional integration was performed.  The results of the spectroscopic
observations of the TTS in MBM12 are summarized in Table~\ref{eqw}.
We list the name of the star; the coordinates for the optical source;
the spectral type; the log of the effective temperature, log$T_{\rm eff}$,
assuming luminosity class V and using the
spectral type-effective temperature relation of \cite{dej87};
apparent magnitude, $V$, taken from the Guide Star Catalog\footnote{Magnitudes
for the sources indicated with a ``:'' are estimated from the
digitized sky survey images};
the equivalent width of H$\alpha$, W(H$\alpha$); both the low and
high resolution (when available) measurements of W(Li);
the veiling corrected W(Li)
(cf. Strom et al. 1989); and the derived lithium abundance based
on the non-LTE curves of growth
of Pavlenko \& Magazz\`u (1996) assuming log$g$=4.5.
The estimated error for the low-resolution W(Li) measurements
is $\sim$ $\pm$ 90 m\AA\ based on the correlation with the three
stars for which we have high
resolution measurements.

\begin{table*}
\caption{H$\alpha$\ and \ion{Li}{I} 6708\AA\ equivalent widths for the stars in MBM12
in which lithium was detected.}
\label{eqw}
\begin{tabular}{@{}c@{\hspace{5pt}}c@{\hspace{5pt}}c@{\hspace{5pt}}c@{\hspace{5pt}}c@{\hspace{5pt}}c@{\hspace{5pt}}c@{\hspace{5pt}}c@{\hspace{5pt}}c@{\hspace{5pt}}c@{\hspace{5pt}}c@{}}
\hline
 Star          & RA         & Dec       & SpT & log$T_{\rm eff}$ & $V$ & W(H$\alpha$)$^{\rm a}$ & W(Li) & W(Li)~hi-res & W(Li) deveiled & logN(Li) \\
               & [2000]     & [2000]    &     &           & [mag] & [\AA]   & [m\AA] & [m\AA] & [m\AA] & \\
\hline
\multicolumn{11}{c}{Main sequence stars} \\
\hline
HD~17332       & 02:47:27.3 & +19:22:24 & G1V & 3.769 &  6.8  &    2.96 & 190 & \nodata & \nodata & 3.2 \\
RXJ0255.3+1915 & 02:55:16.5 & +19:15:02 & F9  & 3.785 & 10.4  &    3.75 & 170 & \nodata & & 3.3 \\
\hline
\multicolumn{11}{c}{T~Tauri stars} \\
\hline
RXJ0255.4+2005 & 02:55:25.7 & +20:04:53 & K6  & 3.631 & 12.2  & $-$1.26  & 380 & $440\pm10$ & 462 & 2.7 \\
LkH$\alpha$262 & 02:56:07.9 & +20:03:25 & M0  & 3.584 & 14.6 & $-$32.1 & 290 & \nodata & 412 & 2.0 \\
LkH$\alpha$263 & 02:56:08.4 & +20:03:39 & M4  & 3.517 & 14.6 & $-$32.9 & 380 & \nodata & 543 & 1.6 \\
LkH$\alpha$264 & 02:56:37.5 & +20:05:38 & K5  & 3.644 & 12.5  & $-$58.9 & 490 & $510\pm20$ & 836 & 3.8 \\
E02553+2018     & 02:58:11.2 & +20:30:04 & K4  & 3.657 & 12.3  & $-$1.6  & 620 & $475^{b}$ & 499 & 3.1 \\
RXJ0258.3+1947 & 02:58:15.9 & +19:47:17 & M5  & 3.501 & 15.0: & $-$24.5 & 580 & \nodata & 783 & 1.8 \\
S18            & 03:02:21.1 & +17:10:35 & M3  & 3.532 & 13.5  & $-$79.0 & 310 & \nodata & 552 & 1.8 \\
RXJ0306.5+1921 & 03:06:33.1   & +19:21:52 & K1  & 3.698 & 11.4  & filled  & 350 & \nodata & \nodata & 3.1 \\
\hline
\end{tabular}

$^a$ A negative sign denotes emission.
$^b$ The high resolution measurement of W(Li) for
this star is taken from Mart\'\i n et al.~(1994).
         
\end{table*}

\begin{figure*}
\vspace{0cm}
\hbox{\hspace{0cm}\epsfxsize=8.9cm \epsfbox{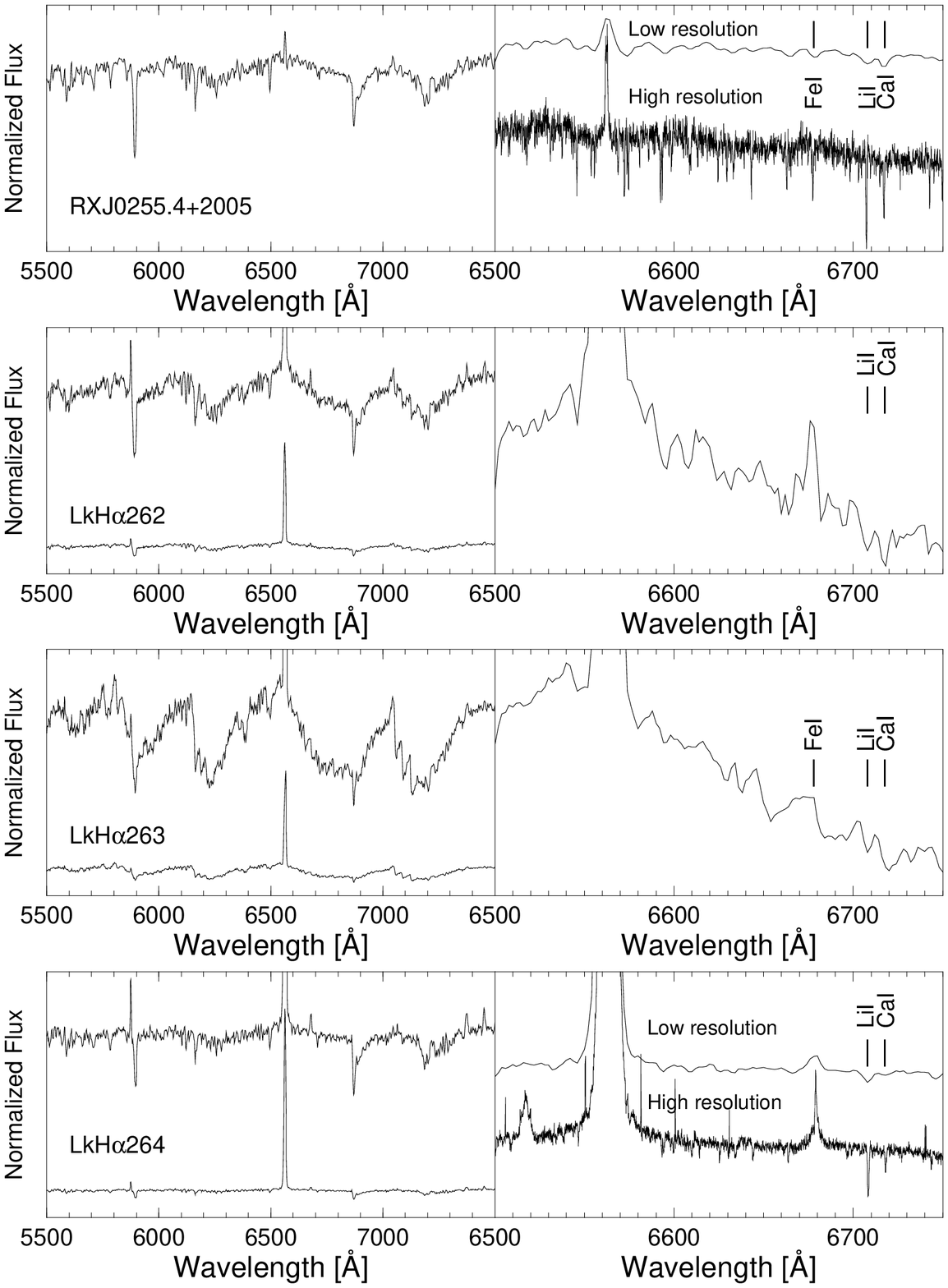}
\epsfxsize=8.9cm \epsfbox{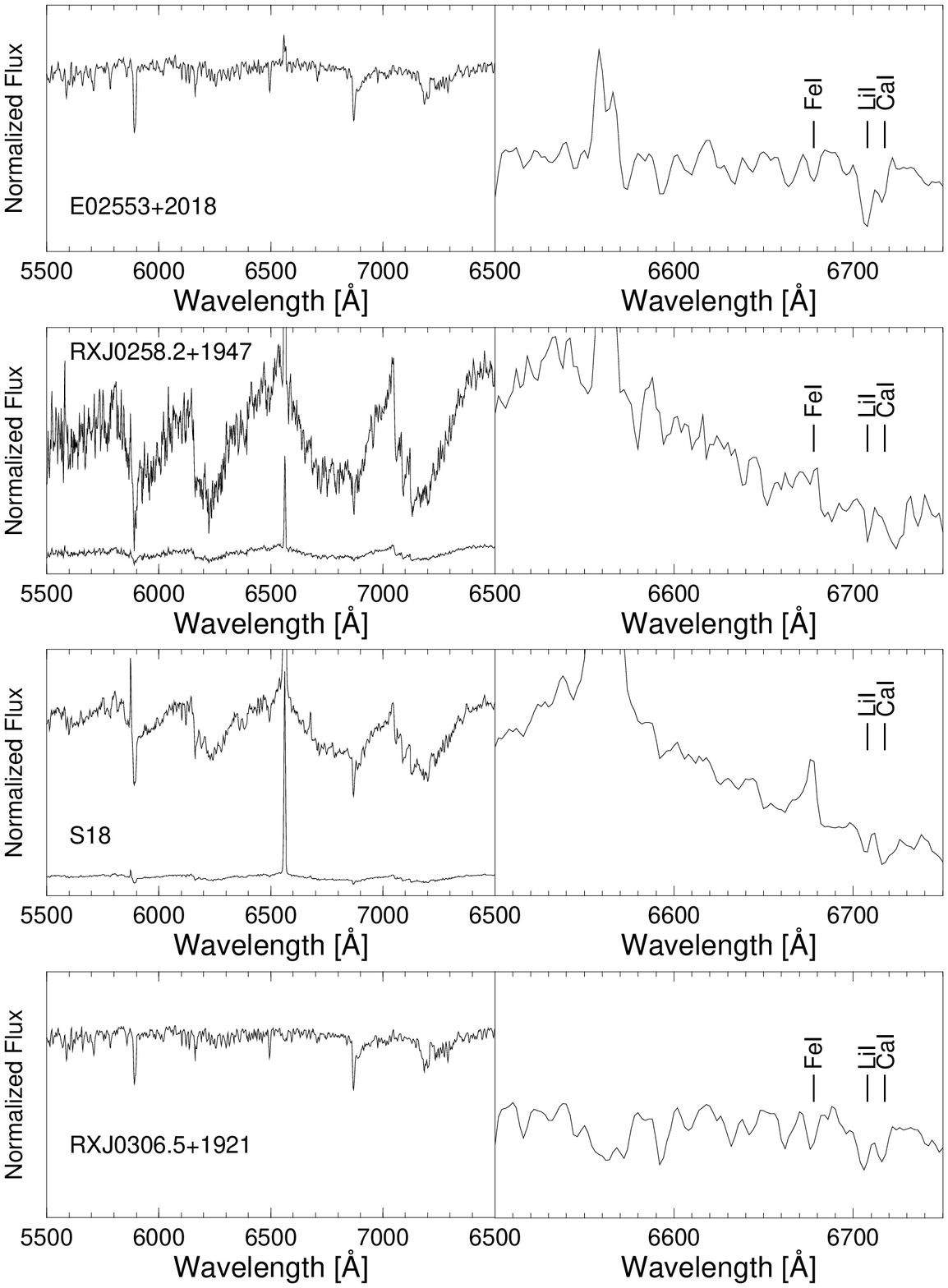}}
\vspace{0cm}
\caption{Spectra are displayed for the 8 currently known TTS in MBM12.}
\label{ttsspectra}
\end{figure*}

\begin{figure}
\resizebox{\hsize}{!}{\includegraphics{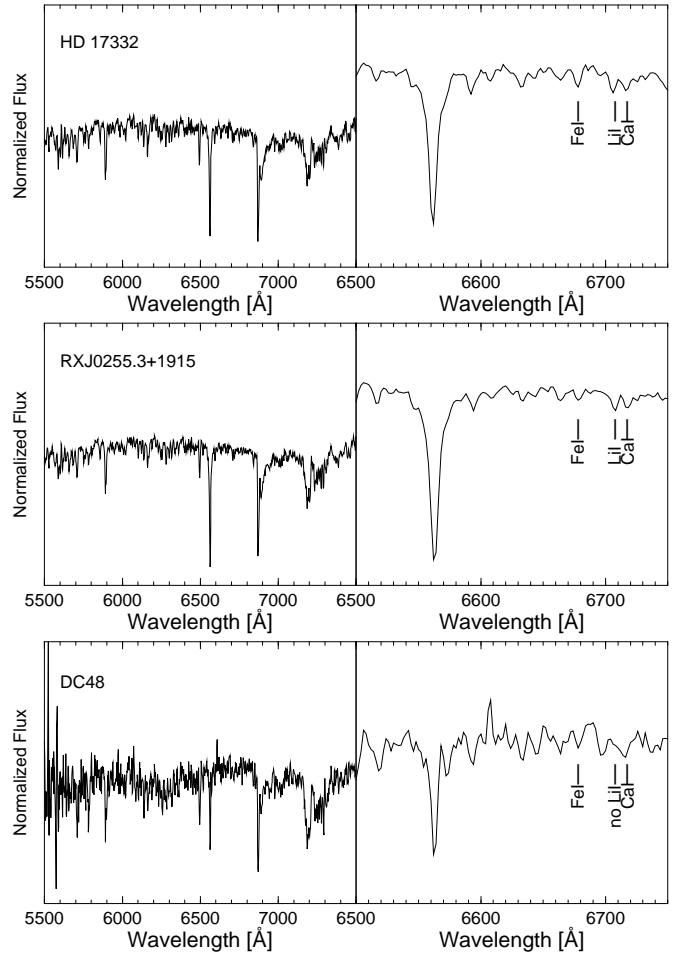}}
\caption{Spectra of the young main sequence stars, HD~17332 and
RXJ0255.3+1915, which have not yet depleted their lithium are displayed.
We also show the spectrum of the highly reddened ($A_{\rm v}$ $\sim$ 8.4--8.9)
G9 star, DC48, which is located behind MBM12.}
\label{zams}
\end{figure}

The optical spectra
of the TTS in MBM12 are displayed in Fig.~\ref{ttsspectra}.
The stars which show strong H$\alpha$ emission are also scaled by an
appropriate factor to display the emission line.
The spectra of the two stars we classify as young main sequence stars
which still show lithium are displayed in Fig.~\ref{zams}.

\begin{figure}
\resizebox{\hsize}{!}{\includegraphics{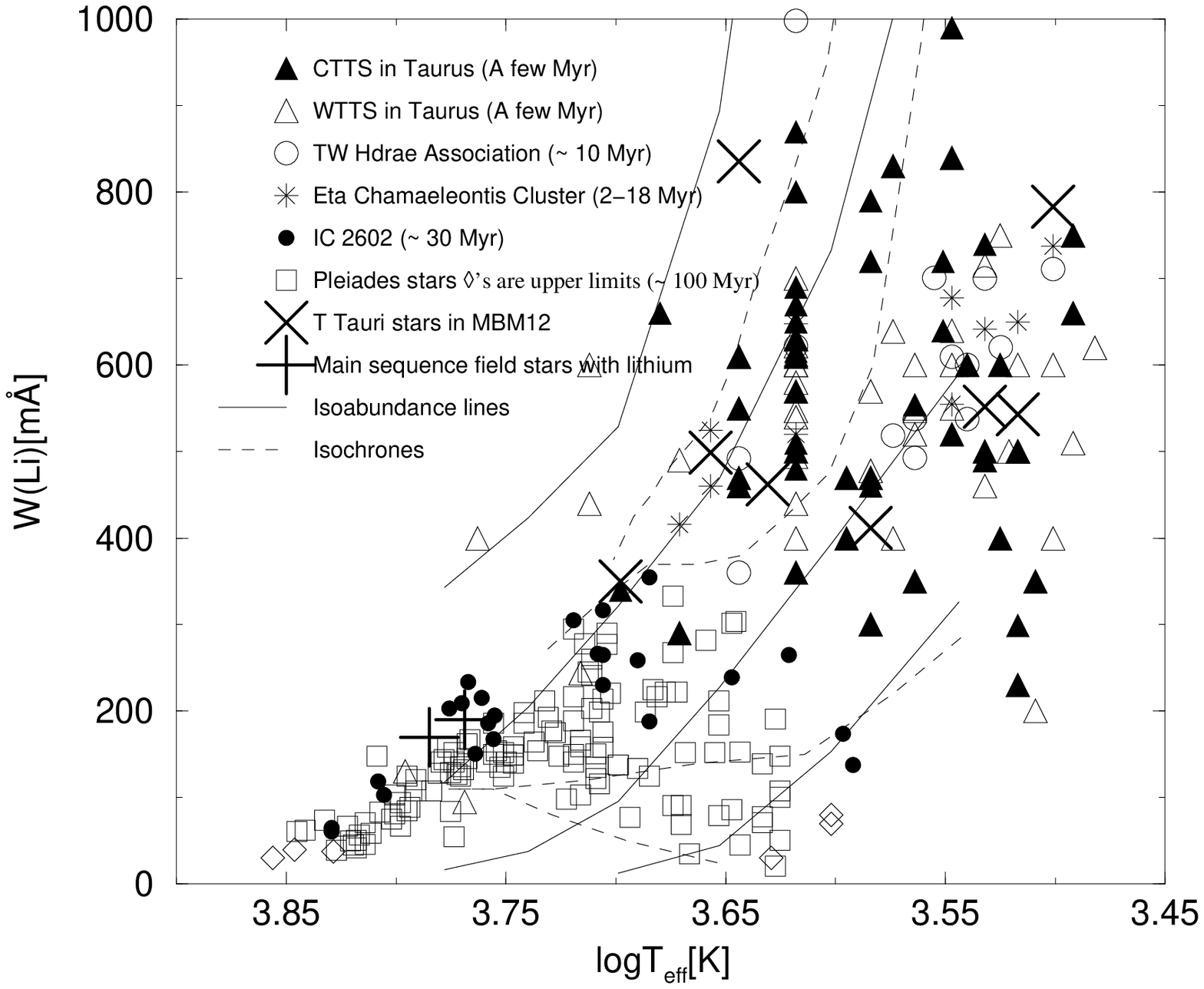}}
\caption{The stars in which we detected lithium are displayed in
an W(Li) v. $T_{\rm eff}$\ diagram.  For comparison, we also
display stars from Taurus
($\sim$ a few Myr; Strom et al. 1989; 
Basri et al. 1991; Patterer et al. 1993; Marcy et al. 1994;
Mart\'\i n et al. 1994), the TW Hydrae Association ($\sim$ 10 Myr;
Webb et al. 1999), the $\eta$ Chamaeleontis Cluster ($\sim$ 2--18 Myr;
Mamajek et al. 1999), IC~2602 ($\sim$ 30 Myr; Randich et al.~1997)
and the Pleiades ($\sim$ 100 Myr; Soderblom et al. 1993).
The solid lines are logN(Li) = 4,3,2,1 isoabundance lines
(from top to bottom, respectively) for the non-LTE curves of
growth of Pavlenko \& Magazz\`u (1996) for log$g$=4.5 stars and
the dashed lines are the 1, 10, 30, 100 Myr isochrones (from top
to bottom, respectively) for the non-rotating
lithium depletion model of Pinsonneault et al.~(1990) normalized to
logN(Li) = 3.3 on a scale where logN(H) = 12.  The location
of the MBM12 TTS in the diagram relative to that of the stars
in the other nearby young clusters indicates they are
$\leq$ 10 Myr.}
\label{lithtemp}
\end{figure}

In addition to confirming that the star S18 is a CTTS with strong
H$\alpha$\ emission and \ion{Li}{I} absorption, we identified 3 previously
unknown TTS in MBM12.  In order to estimate the relative age of the MBM12
stars with lithium we plot them in an
W(Li) vs.~T$_{\rm eff}$\ diagram (Fig.~\ref{lithtemp}) along with stars
from Taurus (age $\sim$~a few Myr), the TW Hydrae Association
(age $\sim$ 10 Myr), the $\eta$ Chamaeleontis Cluster (age $\sim$ 2--18 Myr),
IC~2602 (age $\sim$ 30 Myr), and the Pleiades (age $\sim$ 100 Myr).
In addition, we plot isoabundance lines for the non-LTE curves of
growth of Pavlenko \& Magazz\`u (1996) for log$g$=4.5 stars and
the isochrones for the non-rotating lithium depletion model
of Pinsonneault et al.~(1990).  The positions of the
MBM12 stars in the diagram indicates they are young objects with ages
much less than that of the Pleiades or IC~2602.  Although the
relative ages between the stars in MBM12, the TW Hydrae Association,
and the $\eta$ Chamaeleontis Cluster, cannot be discerned in
Fig.~\ref{lithtemp}, since most of the TTS in MBM12 are
CTTS which are still associated with their parent molecular cloud
the TTS in MBM12 must be younger than those in the TW Hydrae Association
or the $\eta$ Chamaeleontis Cluster which are comprised mainly of WTTS
not associated with any molecular cloud (i.e., the TTS
in MBM12 have ages $<$ 10~Myr).

Although the two F and G spectral type stars in
which we detected lithium (HD~17332 and RXJ0255.3+1915) are located above
the Pleiades in the W(Li) vs.~T$_{\rm eff}$\ diagram,
since they both show H$\alpha$ absorption stronger than any
similar spectral type stars in IC~2602 (e.g., Randich et al.~1997),
they are probably older than 30 Myr.  Thus, we list them as main
sequence stars in Table~\ref{eqw}.
Covino et al. (1997) have shown that low-resolution spectra tend to
overestimate W(Li) in intermediate spectral types, therefore we probably
over estimated the W(Li) for these two stars.

The TTS in MBM12 are clearly
lithium-rich relative to the stars in the Pleiades.  However, current
age dependent stellar population models predict that there should be
a population of young stars with ages $<$ 150 Myr distributed across the sky.
Therefore we compare the density of young X-ray sources detected
in MBM12 with the age dependent stellar population model of Guillout et
al.~(1996) to find out if we are really seeing an excess of young X-ray sources
in the direction of MBM12.  In the galactic latitude range of
40$^{\circ}$ $>$ $\vert$b$\vert$ $>$ 30$^{\circ}$ Guillout et al.~(1996)
predict there should be 0.6--1.0 stars~deg$^{-2}$ and
0.14--0.20 stars~deg$^{-2}$ above
an X-ray count rate threshold of 0.0013 cts s$^{-1}$\ and 0.03 cts s$^{-1}$,
respectively, which have ages $<$ 150 Myr.  Therefore,
we expect to detect $\sim$ 1.9--3.1 young stars in this age group in the
{\it ROSAT} pointed observation and 3.5--5 stars in this age group in the RASS
observation.  Since we observed several young stars which probably
have ages $<$ 150 Myr but are not associated with MBM12 (i.e., the two
intermediate spectral type stars which have not yet depleted their
lithium and the 3 Me and 3 Ke stars listed in Table~4 which have depleted
their lithium but show H$\alpha$ emission), there are a sufficient number
of X-ray active stars in this region to account for the numbers predicted
by Guillout et al.~(1996).  Therefore, the TTS we observe
represent an excess of X-ray active young stars associated with MBM12.

In addition to the X-ray selected T~Tauri star candidates, we also
observed the reddest star from a list of stars compiled by Duerr \& Craine
(1982b) which are along the line of sight to MBM12 and have V-I colors
redder than 2.5~mag.
The optical spectrum of this star, which we will call DC48, indicates
it is a G9 star.  Since Duerr \& Craine (1982b) measured $V$ =  18.7 and
V-I = 5.6~mag, it corresponds to a main sequence star
with $A_{\rm v}$ $\sim$ 8.9 mag at a distance of $\sim$ 63~pc or a giant star
with $A_{\rm v}$ $\sim$ 8.4 mag at a distance of $\sim$ 950~pc.
The spectrum of the highly reddened ($A_{\rm v}$ $\sim$ 8.4--8.9) G9 star,
DC48, is displayed in Fig.~\ref{zams}.

\section{X-ray variability of the TTS}
\label{xrayvar}

\begin{figure}
\hspace{0.0cm}\psfig{figure=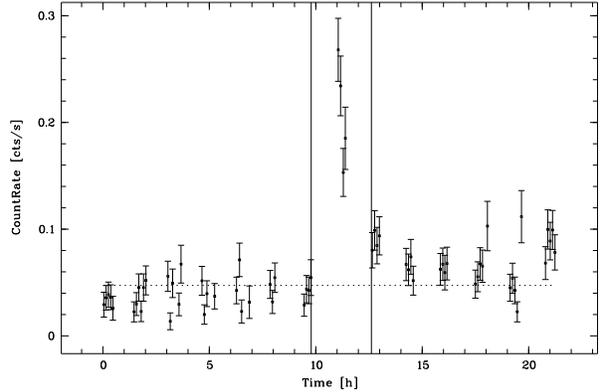,width=6.0cm,angle=-90}
\caption{An X-ray light curve is displayed for the star, RXJ0255.4+2005,
that was found to flare during the {\it ROSAT} pointed observation.
The $\sim$~25~ks exposure of the star is displayed in 400~second bins.
The vertical lines mark the divisions between the pre-flare, flare,
and post-flare phases used for the X-ray spectral fits.}
\label{lc}
\end{figure}

\begin{figure*}
\vspace{0cm}
\psfig{figure=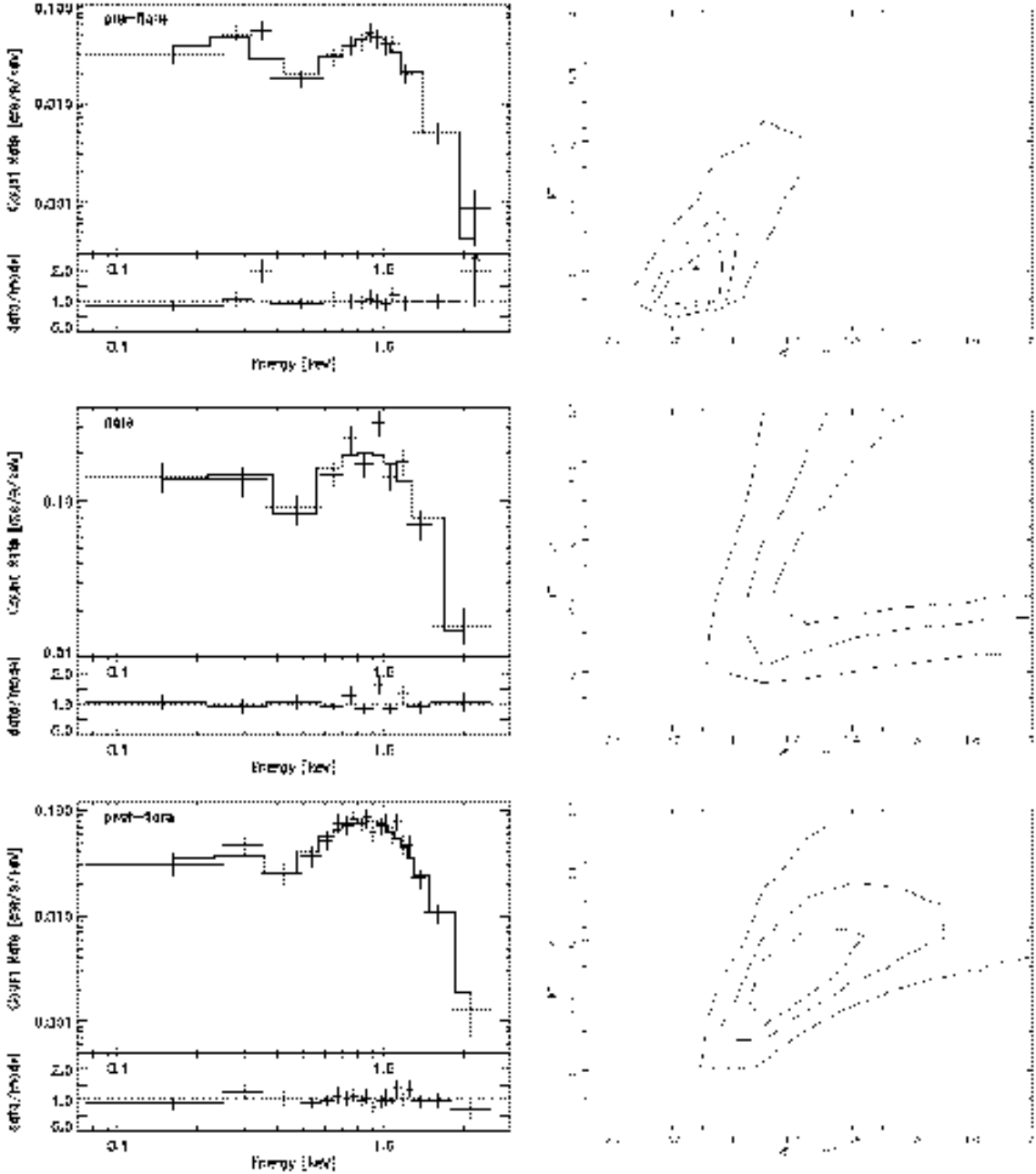,width=17.5cm}\hspace{0cm}
\vspace{0cm}
\caption{The X-ray spectra and the best-fit 2 temperature Raymond-Smith model
are displayed for the pre-flare, flare, and post-flare phases
of the {\it ROSAT} PSPC observation of the weak-line T~Tauri star
RXJ0255.4+2005.  The corresponding $\chi^2$\ plots indicate that both
temperature components increased during the flare.}
\label{xspecflare}
\end{figure*}

We tested all of the TTS for X-ray variability using the methods described
in \cite{ham99}.  The only T Tauri Star which showed X-ray variability
is the newly identified star RXJ0255.5+2005 that was
detected both in the RASS and in the {\it ROSAT} pointed observation and
flared during the pointed observation
(see the light curve displayed in Fig.~\ref{lc}).
The peak X-ray count rate during the flare increased by more than
a factor of 6 from the pre-flare count rate.
Although we do no have a sufficient number of counts
($\sim$~1000 counts for the non-flare phase and $\sim$~500 counts for the
flare phase) for a detailed study of the evolution of the
coronal temperature during the flare,
we performed a rough spectral fit using a 2 temperature Raymond-Smith
model (Raymond \& Smith 1977) including a photoelectric absorption term
using the Morrison \& McCammon (1983) cross sections.
We fit the data for 3 time intervals: the pre-flare phase, the flare,
and the post-flare phase Fig.~\ref{xspecflare}.
Both temperature components increased during the flare 
and remained high throughout the post-flare phase.
The parameters derived from the X-ray spectral fits are listed
in Table~\ref{xraylum} (see Sect.~\ref{xlf} for a description of the
Table columns).  Since the two temperature components are not well
constrained by the spectral fit during the flare, these estimates should
be viewed as a lower limits.  The results of the spectral fits are
consistent with the type of coronal heating seen in high signal-to-noise
X-ray spectra of other flaring WTTS (e.g., Tsuboi et al.~1998).

\section{The X-ray luminosity function}
\label{xlf}

Although our X-ray spectra do not have sufficiently high signal-to-noise
for a detailed comparison of X-ray spectral
models, we performed a spectral fit using a 2 temperature
Raymond-Smith model including a photoelectric absorption
term as described in Sect.~\ref{xrayvar} for the sources with at
least 100 counts.  For the sources with fewer than 100 counts we
calculate the X-ray flux using an X-ray count rate to flux conversion factor
of 1.1 $\times$ 10$^{-11}$ erg~cts$^{-1}$~cm$^{-2}$ which is the mean
conversion factor derived from the spectra for which we performed
spectral fits.  We list the total {\it ROSAT} broad band (0.08--2.0~keV)
counts and count rates for the TTS in MBM12 and the derived
interstellar+circumstellar absorption cross sections and plasma
temperatures for the spectra in which we performed spectral fits
in Table~\ref{xraylum}.  The X-ray luminosities
assume a distance of 65~pc.
Since the binary LkH$\alpha$262/263 was not spatially resolved with
the PSPC we fit the combined X-ray spectrum to estimate the combined
X-ray luminosity but we divide that value in half to generate the X-ray
luminosity function.

\begin{table*}
\caption{X-ray parameters in the 0.08--2.0 keV band for the TTS in MBM12}
\label{xraylum}
\begin{tabular}{@{}l@{\hspace{9pt}}c@{\hspace{5pt}}c@{\hspace{9pt}}c@{\hspace{5pt}}c@{\hspace{9pt}}c@{\hspace{5pt}}c@{\hspace{5pt}}c@{\hspace{5pt}}c@{\hspace{1pt}}c@{\hspace{1pt}}c@{}}
\hline
 Star          & \multicolumn{2}{c}{RASS} & \multicolumn{2}{c}{Pointed Obs.} & \\
\cline{2-3} \cline{4-5}
               & Counts  &  Rate  &  Counts & Rate & N$_{\rm H}$/10$^{21}$ & $kT_{\rm high}$  & $kT_{\rm low}$ &  $\chi ^2$/dof  & $f_{\rm x}$/10$^{-13}$ & log$L_{\rm x}$ \\
               &                & [cts s$^{-1}$] & & [cts s$^{-1}$]  & [cm$^{-2}$] & [keV] & [keV] &  & [erg~s$^{-1}$~cm$^{-2}$] &[erg~s$^{-1}$] \\
\cline{2-3} \cline{4-5}

\hline
RXJ0255.4+2005 \\
`` '' total          & $17\pm5$ & $0.04\pm0.01$ & $1719\pm52$ & $0.067\pm0.002$  & 0.86  & 0.89 & 0.08   & 39.7/34 & 7.81 & 29.60 \\ 
``  ''  pre-flare      & \nodata  & \nodata       & $524\pm29$  & $0.043\pm0.002$  & 0.49  & 0.88 & 0.11   & 9.2/9 & 4.12 & 29.32 \\    
``  ''    flare      & \nodata  & \nodata       & $368\pm22$  & $0.22\pm0.01$  & 0.18  & 1.69 & 0.30   & 9.5/6 & 20.1 & 30.01 \\    
``  '' post-flare      & \nodata  & \nodata       & $781\pm35$ & $0.071\pm0.003$  & 0.27  & 1.14 & 0.24   & 10.7/15 & 6.96 & 29.55 \\
LkH$\alpha$262/263   & \nodata  & \nodata       & $340\pm21$  & $0.016\pm0.001$  & 2.65  & 0.97 & 0.16   & 6.29/8  & 2.01 & 29.01 \\ 
LkH$\alpha$264       & \nodata  & \nodata       & $89\pm12$   & $0.0043\pm0.0006$&\nodata&\nodata&\nodata& \nodata & 0.47 & 28.39 \\ 
E02553+2018          & $9\pm4$  & $0.04\pm0.02$ & $575\pm34$  & $0.036\pm0.002$  & 8.21  & 1.05 & 0.10   & 16.5/16 & 5.03 & 29.41 \\ 
RXJ0258.3+1947       & \nodata  & \nodata       & $182\pm16$  & $0.0079\pm0.0007$& 8.42  & 1.14 & 0.13   & 2.63/2  & 0.86 & 28.64 \\ 
S18                  & $6\pm3$  & $0.03\pm0.01$ & \nodata     & \nodata          &\nodata&\nodata&\nodata& \nodata & 3.30 & 29.22 \\ 
RXJ0306.5+1921       & $12\pm4$ & $0.03\pm0.01$ & \nodata     &\nodata           &\nodata&\nodata&\nodata& \nodata & 3.30 & 29.22 \\ 
\hline
\end{tabular}

\end{table*}

\begin{figure}
\hspace{0.0cm}\psfig{figure=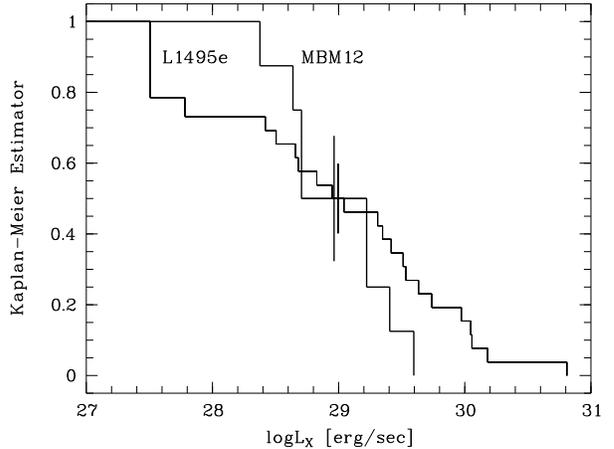,width=7.0cm,angle=-90}
\caption{The X-ray luminosity function for the TTS in MBM12
is displayed with a similar X-ray luminosity function for the
TTS in the L1495E cloud in Taurus.  Distances of 65~pc and 140~pc
are assumed for the X-ray luminosity functions of MBM12 and L1495E,
respectively.  The error bars shown are the largest for each data set.}
\label{lumfunc}
\end{figure}

In order to compare the derived X-ray luminosity
function for the TTS in MBM12 with other flux limited X-ray luminosity
functions we used the ASURV Rev.~1.2 package (Isobe \& Feigelson 1990;
LaValley et al.~1992), which
implements the methods presented in Feigelson \& Nelson (1985).
Although the currently known TTS in MBM12 are all X-ray detected, the
luminosity functions of other, more distant, star forming regions
include upper limits. 
The derived X-ray luminosity function is displayed in Fig.~\ref{lumfunc}
with the X-ray luminosity function for the TTS in the L1495E cloud in
Taurus which (like MBM12) was observed in a deep (33 ks) {\it ROSAT}
PSPC pointed observation (Strom \& Strom 1994).  The {\it ROSAT} pointed
observation of L1495E is $\sim$ 20 times more sensitive than previous
observations with the {\it Einstein} satellite.  Strom \& Strom (1994)
used this observation to show that the X-ray
luminosity of TTS extends to fainter luminosities than were observed
with {Einstein}.  We have re-reduced the pointed observation of L1495E
in a way analogous to that of MBM12.  The X-ray luminosity function
we derive for L1495E (1) includes only the K and M spectral type TTS,
(2) includes 6 upper limits, (3) assumes an X-ray to optical flux
conversion factor of 1.1~$\times$~10$^{-11}$~erg~cts$^{-1}$~cm$^{-2}$, and
(4) assumes a distance of 140 pc.  The X-ray luminosity
functions in MBM12 and L1495E agree
well: in MBM12 the log$L_{\rm x~mean}$ = 29.0$\pm$0.1 erg s$^{-1}$and
log$L_{\rm x~median}$ = 28.7 erg s$^{-1}$; in L1495E log$L_{\rm x~mean}$ =
28.9$\pm$0.2 erg s$^{-1}$ and log$L_{\rm x~median}$ = 28.9 erg s$^{-1}$.
However, we note that the MBM12 X-ray luminosity function has a lower
high-luminosity limit and a higher low-luminosity limit than the
L1495E X-ray luminosity function.  Therefore, although the pointed observation
of MBM12 is more sensitive than the pointed observation L1495E (because
MBM12 is much closer) our follow-up observations of the TTS in MBM12
may be incomplete for sources fainter than $V$ $\sim$ 15.5~mag.  
In addition, since we know that
one of our X-ray sources, S18, is detected but below our threshold
for follow-up observations, there may be other, fainter, X-ray
emitting TTS in MBM12 with spectral types later than $\sim$ M2
(i.e., the spectral type of S18) that will be discovered in more
sensitive follow-up observations.
The discrepancy at the high luminosity end of the X-ray luminosity function
may also be explained if the distance to the TTS in MBM12 is larger than
65~pc.  Although an increased distance is allowed by the recent
{\it Hipparcos} results it should be confirmed with further observations.

\section{Conclusions}
\label{conclusions}
 
Although MBM12 is not a prolific star-forming cloud when compared to
nearby giant molecular clouds it is the
nearest star-forming cloud to the sun and offers a unique opportunity
to study the star-formation process within a molecular cloud at high
sensitivity.  We have presented follow-up observations of X-ray stars
identified in the region of the MBM12 complex.  These observations have
doubled the number of confirmed TTS in this region. 
Since the {\it ROSAT} PSPC pointed observation of the central region
of the cloud was sensitive enough to detect all of the previously
known TTS in the cloud, we believe the list of 5 CTTS and 3 WTTS
presented in Table~\ref{eqw} to be a nearly complete census of
the TTS in MBM12 for spectral types earlier than $\sim$~M2. 
Assuming a mean mass of $\sim$ 0.6~M$_{\odot}$ for the 8 currently known
TTS in MBM12 and a cloud mass of 30--200~M$_{\odot}$ (Pound et al.~1990;
Zimmermann \& Ungerechts 1990) the star-formation
efficiency of MBM12 is
$\sim$~2--24\%.  Since the currently known TTS population in MBM12 is
incomplete only for the lower mass objects, unless there are a huge number
of these objects yet to be discovered in the cloud, this estimate of the
star-formation efficiency will not change significantly.  Although there
is still a large uncertainty in the mass of the cloud the estimated
star-formation efficiencies are consistent with that expected from
clouds with masses on the order of 100~M$_{\odot}$ (Elmegreen \& Efremov 1997).
By comparing the strengths of the H$\alpha$ emission and
\ion{Li}{I} $\lambda$6708~\AA\ absorption lines of the TTS in MBM12
with those found in other young clusters, we place an upper limit on the
age of the stars in MBM12 $\sim$ 10~Myr.

By comparing the X-ray luminosity function of
the TTS in MBM12 with that of the TTS in L1495E we predict
that there are more young, low-mass, stars to be discovered
in MBM12 and the assumed distance to the cloud may have to be
increased.  Although this prediction agrees with the recently
revised distance estimate to the cloud ($\sim$ $65\pm35$~pc) based on 
results of the {\it Hipparcos} satellite, it should be confirmed with future
observations.  

We have also identified a reddened G9 star behind the cloud with
$A_{\rm v}$ $\sim$ 8.4--8.9 mag.  Therefore, there are at least two lines of
sight through the cloud that show larger
extinctions ($A_{\rm v}$ $>$ 5~mag) than previously thought for this cloud. 
This higher extinction explains why MBM12 is capable of star-formation
while most other high-latitude cloud are not.

\begin{acknowledgements}

We wish to thank Patrick Guillout for helpful discussions
about the expected population of young X-ray active stars located
at high galactic latitude and Loris Magnani for insightful comments
concerning this paper.  We also thank an anonymous referee for
suggestions which enable us to put firmer constraints on the age
of the TTS in MBM12.
The {\it ROSAT} project is supported by the Max-Planck-Gesellschaft and
Germany's federal government (BMBF/DLR).
TH is grateful for a stipendium from the Max-Planck-Gesellschaft
for support of this research.  RN acknowledges
a grant from the Deutsche Forschungsgemeinschaft (DFG
Schwerpunktprogramm ``Physics of star formation'')

\end{acknowledgements}


\begin{thebibliography}{}

\bibitem[1999]{bal99} Ballesteros-Paredes J., Hartmann L.,
V\'azquez-Semadeni E., 1999, ApJ, in press

\bibitem[Basri et al. 1991]{bas91} Basri G., Mart\'\i n E.L.,
Bertout C., 1991, A\&A 252, 625

\bibitem[1992]{bha95} Bhatt H. C., Jain S. K., 1992, A\&A 257, 57 

\bibitem[Briel \& Pfeffermann (1995)]{bri95} Briel U.G., Pfeffermann
E., 1995, in Proc. SPIE Vol. 2518, {\it EUV, X-ray, \& Gamma-Ray
Instrumentation for Astronomy VI}, eds. O.H. Siegmund, J.V. Vallerga, 120

\bibitem[Caillault et al. 1995]{cai95} Caillault J.-P., Magnani L.,
Fryer C., 1995, ApJ 441, 261

\bibitem[Covino et al. 1997]{cov97} Covino E., Alcal\'a J.M., Allain S.,
et al. 1997, A\&A 328, 187

\bibitem[Damiani et al. 1995]{dam95} Damiani F., Micela G., Sciortino S.,
Harnden F.R., Jr., 1995, ApJ 446, 331

\bibitem[de Jager \& Nieuwenhuijzen (1987)]{dej87} de Jager C.,
Nieuwenhuijzen H., 1987, ApJ 177, 217

\bibitem[Downes \& Keyes 1988]{dow88} Downes R.A., Keyes C.D., 1988,
AJ 96, 1988

\bibitem[Duerr \& Craine 1982a]{due82a} Duerr R., Craine E.R., 1982a, AJ 87, 408

\bibitem[Duerr \& Craine 1982b]{due82b} Duerr R., Craine E.R., 1982b, PASP 94, 567 

\bibitem[Elmegreen 1993]{elm93}Elmegreen B.G., 1993, in {\it Protostars and Planets III}, eds. E. Levy, J. Lunine (University of Arizona Press), 97

\bibitem[]{} Elmegreen B.G., Efremov Y.N., 1997, ApJ 480, 235

\bibitem[Feigelson 1996]{fei96} Feigelson E., 1996, ApJ 468, 306

\bibitem[]{} Feigelson E., Nelson P.I., 1985, ApJ 293, 192

\bibitem[Fern\'andez et al. 1995]{fer95} Fern\'andez M., Ortiz E., Eiroa C.,
Miranda L.F., 1995, A\&AS 114, 439

\bibitem[Gameiro et al. 1993]{gam93} Gameiro J.F., Lagor M.T.V., Lima N.M.,
Cameron A.C., 1993, MNRAS 261, 11

\bibitem[Gioia et al. 1990]{gio90} Gioia M., Maccacaro T., Schild R.E.,
et al., 1990, ApJS 72, 567

\bibitem[Guillout et al. 1996]{gui96} Guillout P., Haywood M., Motch C.,
Robin A.C., 1996, A\&A 316, 89

\bibitem[Hambaryan et al.~(1999)]{ham99} Hambaryan V., Neuh\"auser R., Stelzer B., 1999, A\&A 345, 121

\bibitem[Hearty et al. 1999]{hea99} Hearty T., Magnani L., Caillault J.-P.,
et al., 1999, A\&A 341, 163
 
\bibitem[Herbig 1977]{her77} Herbig G.H., 1977, ApJ 214, 747

\bibitem[Herbig \& Bell 1988]{her88} Herbig G.H., Bell K.R., 1988,
Lick Observatory Bulletin 1111, 1
 
\bibitem[Hobbs et al.~(1986)]{hob86} Hobbs L.M., Blitz L.,
Magnani L., 1986, ApJ 306, L109

\bibitem[Hobbs et al. (1988)]{hob88} Hobbs L.M., Blitz L., Penprase B.E.,
Magnani L., Welty D.E., 1988, ApJ 327, 356 

\bibitem[]{} Isobe T., Feigelson E., 1990, BAAS 22, 917

\bibitem[Kastner et al. 1997]{kas97} Kastner J.H., Zuckerman B.,
Weintraub D.A., Forveille T., 1997, Science 277, 67

\bibitem[LaValley, Isobe, \& Feigelson 1992]{lav92} LaValley M., Isobe T.,
Feigelson E., 1992, BAAS 1, 245

\bibitem[L\'epine \& Duvert 1994]{lep94} L\'epine J.R.D., Duvert G., 1994,
A\&A 286, 60

\bibitem[Lynds 1962]{lyn62} Lynds B.T., 1962, ApJS 7, 1

\bibitem[Magnani et al. 1985]{mag85} Magnani L., Blitz L., Mundy L., 1985,
ApJ 295, 402

\bibitem[Magnani et al. 1995]{mag95} Magnani L., Caillault J.-P.,
Buchalter A., Beichman C.A., 1995, ApJS 96, 159

\bibitem[]{} Mamajek E., Lawson W.A., Feigelson E.D., 1999, ApJ 516, 77

\bibitem[Marcy et al. 1994]{marbas94} Marcy G.W., Basri G., Graham J.R., 1994,
ApJ 428, L57

\bibitem[Mart\'\i n et al. 1994]{mar94} Mart\'\i n E.L., Rebolo R.,
Magazz\`u A., Pavlenko Ya.V. 1994, A\&A 282, 503

\bibitem[]{} Moriarty-Schieven G.H., Andersson B.-G., Wannier P.G., 1997,
ApJ 475, 642

\bibitem[]{} Morrison R., McCammon, D. 1983, ApJ 270, 119

\bibitem[Neuh\"auser et al. 1995]{neu95} Neuh\"auser R.,
Sterzik M.F., Schmitt J.H.M.M, Wichmann R., Krautter J.,
1995, A\&A 297, 391

\bibitem[Patterer et al. 1993]{pat93} Patterer R.J., Ramsey L.,
Huenemoerder D.P., Welty A.D., 1993, AJ 105, 1519

\bibitem[1996]{pav96} Pavlenko Ya. V., Magazz\`u A., 1996, A\&A 311, 961

\bibitem[1990]{pin90} Pinsonneault M. H., Kawaler S. D., Demarque P.,
1990, ApJ 74, 501

\bibitem[Pound, Bania, \& Wilson 1990]{pou90} Pound M.W., Bania T.M.,
Wilson R.W., 1990, ApJ 351, 165

\bibitem[]{} Queloz D., 1994, in IAU symposium 167,Ed. A.G. Davis Philip, p. 221

\bibitem[1997]{ran97} Randich S., Aharpour N., Pallavicini R., Prosser C. F.,
Stauffer J. R., 1997, A\&A 323, 86

\bibitem[]{} Raymond J.C., Smith B.W., 1977, ApJS 35, 419

\bibitem[]{} Soderblom D., Pendelton J., Pallavicini R., 1989, AJ 97, 539

\bibitem[Soderblom et al. 1993]{sod93} Soderblom D.R., Jones B.F.,
Balachandran S., et al., 1993, AJ 106, 1059

et al., 1994, ApJS 91, 625

\bibitem[]{} Stephenson C.B., 1986, ApJ 300, 779

\bibitem[]{} Sterzik M.F., Durisen R., 1995, A\&A 304, L9

\bibitem[Stocke et al. 1991]{sto91} Stocke J.T., Morris S.L.,
Gioia I.M., et al., 1991, ApJS 76, 813

\bibitem[]{} Strom K.M., Strom S.E., 1994, ApJ 424, 237

\bibitem[Strom et al. 1989]{str89} Strom K., Wilkin F., Strom S.,
Seaman R., 1989, AJ 98, 1444

\bibitem[Tr\"umper (1983)]{tru83} Tr\"umper J., 1983, Adv. Space Res. 2, 241

\bibitem[]{} Tsuboi Y., Koyama K., Murakami H., et al., 1998, ApJ 503, 894

\bibitem[1999]{web99} Webb R. A., Zuckerman B., Platais I., et al., 1999,
ApJ 512, L63

\bibitem[]{} Zimmermann T., Ungerechts H., 1990, A\&A 238, 337	

\bibitem[]{} Zuckerman B., Becklin E.E., McLean I.S., Patterson J., 1992,
ApJ 400, 665

\end{thebibliography}
\end{document}